%% file: sample-sigconf.tex
\documentclass[sigconf]{acmart}

\usepackage{booktabs} 

\usepackage{algorithm2e}

\begin{document}
\setcopyright{rightsretained}
\acmConference[SIGIR 2018 eCom]{ACM SIGIR Workshop on eCommerce}{July 2018}{Ann Arbor, Michigan, USA}
\acmYear{2018}
\copyrightyear{2018}

\title{Leveraging Catalog Knowledge Graphs for Query Attribute Identification in E-Commerce Sites}
\author{Suhas Ranganath}
\affiliation{%
  \institution{Walmart Labs}
  }
\email{suhas.ranganath@walmartlabs.com}

\renewcommand{\shortauthors}{S. Ranganath et al.}

\begin{abstract}
Millions of people use online e-commerce platforms to search and buy products. Identifying attributes in a query is a critical component in connecting users to relevant items. However, in many cases, the queries have multiple attributes, and some of them will be in conflict with each other. For example, the query ``maroon 5 dvds'' has two candidate attributes, the color ``maroon'' or the band ``maroon 5'', where only one of the attributes can be present. In this paper, we address the problem of resolving conflicting attributes in e-commerce queries. A challenge in this problem is that knowledge bases like Wikipedia that are used to understand web queries are not focused on the e-commerce domain. E-commerce search engines, however, have access to the catalog which contains detailed information about the items and its attributes. We propose a framework that constructs knowledge graphs from catalog to resolve conflicting attributes in e-commerce queries. Our experiments on real-world queries on e-commerce platforms demonstrate that resolving conflicting attributes by leveraging catalog information significantly improves attribute identification, and also gives out more relevant search results.
\end{abstract}

\maketitle

\input{samplebody-conf}

\bibliographystyle{ACM-Reference-Format}
\bibliography{sample-sigconf}

\end{document}

%% file: samplebody-conf.tex
\section{Introduction}
E-commerce sites are being used by millions of people to buy products in a fast and seamless manner. Users express their buying needs through search queries, and an accurate understanding of the query is necessary to return relevant items ~\cite{Baeza-Yates:2017}. A crucial part of query understanding is to identify attributes inherent in the query ~\cite{li2013unsupervised}. For example, identifying that the query ``maroon 5 dvds'' has product type ``dvd'' helps in returning relevant items.  Research on identifying query attributes in web search explores the use of semantic information ~\cite{li2016deep}, user engagement ~\cite{ren2014heterogeneous} and external knowledge bases ~\cite{wang2015query}. There has been relatively less work in identifying query attributes in the e-commerce domain.  

In many cases, query understanding systems have conflicting candidate attributes for a given query. In the example query, ``maroon 5 dvds'' the candidate attributes are the product type ``dvds'' the color ``maroon'' and the band ``maroon 5''. It is not straightforward for query understanding systems to infer whether the query is referring to the band ``maroon 5'' or the color ``maroon''.   Designing algorithms to resolve conflicting query attributes can help e-commerce search systems to return more relevant items and better satisfy the buying needs of users.

This task faces several challenges. First, queries are short and contain insufficient information for systems to identify attributes. Second, knowledge bases like Wikipedia used to supplement query text in web search ~\cite{hu2009understanding} are not focused on e-commerce domain, and can lead to insufficient and noisy information. Third, e-commerce sites have millions of users, and search algorithms have significant issues of scalability.

E-commerce search systems have access to a catalog which contains the attributes of items sold by the system. Leveraging the e-commerce catalog as a knowledge base to supplement the textual information can help to resolve conflicting query attributes. In the example query ''maroon 5 dvds'', we see from the catalog that there is a significant number of items having the product type as ``dvds'' and have a band attribute whereas very few of them have the color attribute. This indicates that the catalog can provide valuable information which can use to resolve conflicting query attributes. Therefore, in this paper, we propose a framework to model catalog information to better identify attributes in e-commerce queries.

Specifically, we address the following questions: How to model the catalog information to resolve conflicts in query attributes? How to evaluate the impact of the framework on on e-commerce search systems? The primary contributions of the work are
\begin{itemize}
    \item Proposing the problem of resolving conflicts in query attributes for e-commerce queries;
    \item Proposing a framework to model attribute relations from the catalog to identify query attributes in e-commerce queries; and
    \item Presenting evaluations of the utility of catalog information in identifying query attributes on real-world data.
\end{itemize}

The rest of the paper is organized as follows. In Section 2, we describe the proposed framework. In Section 3, we present evaluations of the framework for identifying the query attribute and its impact on ranking relevant items. We conclude in Section 4 along with possible future directions.

\section{The Proposed Framework}
In this section, we present our proposed framework to identify attributes in e-commerce queries. We first describe the notations used and then define the problem statement. We then use the notations to describe various aspects of the framework.

We now present the notations used in the paper. Let $q$ be the query and $\mathcal{A}=\{a_1,a_2,...,a_n\}$ be the set of candidate attributes in the query. Let $\{\mathcal{A}-a_k\}$ be the set of attributes of size $N_a$ without the attribute $a_k$.  The problem can be formally stated as follows: ``\textit{Given the query $q$, attribute $a_k$, and the  attribute set  $\{\mathcal{A}-a_k\}$ determine whether attribute $a_k$ is present in $q$.''}

We next present our framework to identify attribute values for a given query. We explore two sets of metrics to model catalog information to assist in query attribute identification. We next describe two sets of metrics along with mathematical formulations, one related to the presence of the attribute in the query, and the second related to the presence of attribute value in the query.

The first metric set computes $p(m/q)$, the probability of an attribute $m$ being present in the query $q$. This is formulated as 
\begin{align}
& p(m/q) \propto  p(m/n=x)*p(n=x/q) \nonumber\\
&p(m/n=x) =  \frac{N(n=x,m)}{degree(n=x)},
\label{eqn:1}
\end{align}
where $p(m/n=x)$ is the probability of attribute $m$ being present where $n=x$, and $degree(n_x)$ is the number of items in the catalog which have value $x$ for attribute $n$. Among such items, $N(n=x,m)$ is the number of items having values for attribute $m$. For a given attribute $m$, we compute Eq \ref{eqn:1} for all attributes $n \in \{\mathcal{A}-m\}$, resulting in a total of $N_a -1$ feature values. According to Eq \ref{eqn:1}, if the query has attribute value $n=x$, it is more likely to have attribute $m$ if more items in the catalog with $n=x$ also contain attribute $m$. In the query ``maroon 5 dvds'', very few items which have the product type ``dvds'' have values for color, and  Eq \ref{eqn:1} metric gives a lower value for $p(color/q)$, the probability of color attribute being present in the query.

The second metric computes $p(m=l/q)$, the probability of attribute $m$ having a value $l$ for the query $q$. This is formulated as 
\begin{align}
&p(m=l/q) \propto p(m=l/n=x)*p(n=x/q)\nonumber\\
&p(m=l/n=x) = \frac{log(N(n=x,m=l))}{log(degree(n=x))},
\end{align}
where the number of items in the catalog having the value $x$ for a given attribute $n$ is $degree(n=x)$. Among these items, let the number of items having value $l$ for attribute $m$ be denoted by $N(n=x,m=l)$. The score is higher if more number of items having the value $x$ for attribute $n$ also have value $l$ for attribute $m$. We repeat this for all possible attributes $n \in \{\mathcal{A}-m\}$ for a given attribute $n$ resulting in an additional $N_a -1$ feature values. The value set for a given attribute follow a power law distribution where few values are prominent, so we employ log smoothing to make the values linearly distributed.

We integrate the scores derived from the catalog metrics into a feature set. Our metrics are scalable and hence suited for handling large-scale traffic common in an e-commerce site. We use an out of the box classifier on the feature set to determine whether or not the given query has the attribute value.

\section{Evaluation}
We next evaluate our framework with the help of traffic weighted random sample of 20000 queries on Walmart.com. To compare our framework, we use the baseline \textbf{Dict Lookup} which identifies attributes for a query by matching overalapping phrases in the query with terms in the attribute dictionary. This baseline does not address the possible conflicts that can arise between candidate attributes. For the evaluation, we take color as the attribute that has to be predicted for a query and the product type and brand as the attributes whose values are known for the query.

We design two evaluation tasks. The first task assesses the effectiveness of the framework on identifying attributes of a given query. We employ manual labeling by expert annotators for the ground truth and use Precision, Recall, and F1 as the evaluation metrics. The second task assesses the impact of the framework on the ranking relevant items for the given query. We use the orders of the query-item pair for the ground truth and nDCG@20 as the evaluation metric. The evaluation results are illustrated in Table \ref{tab:attribute}.

From the table, we can see that the framework is significantly better in identifying attributes of a given e-commerce query than the baseline across all the Precision, Recall and F1 metrics. The improvement in identifying query attributes is also reflected in showing better ranking results as shown by the lift in nDCG@20. The improvement in both the tasks demonstrates that the catalog can be effectively leveraged as a knowledge base to identify attributes for a given query in a better manner, and the ability of the metric to effectively capture the relevant catalog information.

\begin{table}[]
    \centering
    \begin{tabular}{c c c c c}
    \toprule
        \textbf{}&\textbf{Precision}&\textbf{Recall}&\textbf{F1}&\textbf{nDCG@20}\\
    \midrule
        \textbf{Gain} & +6.48\% & +11.37\% & +8.3\% &  +5.36\%\\
    \bottomrule
    \end{tabular}
    \caption{Performance of the framework on attribute identification and ranking}
    \label{tab:attribute}
\end{table}

\section{Conclusions and Future Work}
In this paper, we address the problem of identifying attributes for queries on e-commerce sites. General purpose knowledge bases used in identifying attributes for web queries are not focused on e-commerce needs.  We design a framework that leverages catalog as a knowledge base to resolve conflicts in query attributes. We evaluate the framework on the set of queries from Walmart.com and demonstrate that it significantly improves results in attribute identification and ranking relevant items for e-commerce queries. Future research directions can include leveraging query catalog interactions and query sequences to design more involved metrics for attribute identification. The utility of catalog in other query understanding tasks such as query reformulation and type-ahead is also an interesting avenue for researchers to explore.